\documentclass[conference]{IEEEtran}
\IEEEoverridecommandlockouts
\usepackage{cite}
\usepackage{amsmath,amssymb,amsfonts}
\usepackage{algorithm}
\usepackage{graphicx}
\usepackage{textcomp}
\usepackage{comment}
\usepackage{xcolor}
\usepackage{subcaption}
\usepackage{amsmath}
\usepackage{amssymb}
\usepackage{gensymb}
\usepackage{multirow}
\usepackage{array}
\usepackage{bm}
\usepackage{subcaption}
\usepackage{stfloats}
\usepackage{tikz}
\usepackage{bm} 
\usepackage{amsmath,amssymb,booktabs,array}
\usepackage[left=1.62cm,right=1.62cm,top=1.9cm]{geometry}
\usepackage{siunitx}
\usepackage{algpseudocode}
\usepackage[nolist,nohyperlinks]{acronym}

\usepackage{tikz}
\usepackage{graphicx}
\usepackage{subcaption}

\usepackage{array}
\usepackage{lipsum}
\usepackage{booktabs}
\usepackage[table]{xcolor}
\usepackage{balance}
\usepackage{tabularx}
\def\BibTeX{{\rm B\kern-.05em{\sc i\kern-.025em b}\kern-.08em
    T\kern-.1667em\lower.7ex\hbox{E}\kern-.125emX}}
\begin{document}

\title{Wind-Resilient Trajectory Optimization for UAV-BS Networks: TD3 for Continuous Service Availability} 

\author{
\IEEEauthorblockN{
Azim Akhtarshenas\IEEEauthorrefmark{1},
German Svistunov\IEEEauthorrefmark{1},
Kuangyu Zheng\IEEEauthorrefmark{3},
and
David López-Pérez\IEEEauthorrefmark{1,2}
}
\normalsize\IEEEauthorblockA{\emph{
\IEEEauthorrefmark{1}Universitat Politècnica de València (UPV), Spain} 
}
\normalsize\IEEEauthorblockA{\emph{
\IEEEauthorrefmark{3}Beihang University, China} 
}
\textit{aakhtar@doctor.upv.es}
\thanks{This research was supported by the Generalitat Valenciana through the CIDEGENT PlaGenT (Grant~CIDEXG/2022/17, Project~iTENTE); the action CNS2023-144333, funded by MCIN/AEI/10.13039/501100011033 and the European Union NextGenerationEU/PRTR; K. Zheng was supported by the China National R\&D Key Program, the Xiongan Innovation Project (SQ2023XAGG0188); the National Natural Science Foundation of China Key Program (U2333201); the Ganwei Program (WZ2026-ZD-16); and the Hebei Science and Technology Program (No.~246Z0303G). }}

\maketitle
\begin{acronym}[AAAAAAAAAAAAAAAAAAAAAAAA]  
 \acro{3GPP}{third generation partnership project}
 \acro{4G}{fourth-generation}
 \acro{5G}{fifth-generation}
 \acro{6G}{sixth-generation}
 \acro{AI}{artificial intelligence}
 \acro{BS}{base station}
 \acro{dB}{decibel}
 \acro{dBi}{decibel isotropic}
 \acro{DRL}{deep reinforcement learning}
 \acro{FR}{Frequency Range}
 \acro{FSO}{free-space optical}
 \acro{HAPS}{high-altitude platform station}
 \acro{IoT}{Internet of things}
 \acro{IoS}{Internet-of-ships}
 \acro{LEO}{low Earth orbit}
 \acro{LoS}{line-of-sight}
 \acro{MCN}{maritime communication network}
 \acro{MDP}{Markov decision process}
 \acro{ML}{machine learning}
 \acro{NLoS}{non-line-of-sight}
 \acro{NTN}{non-terrestrial network}
 \acro{PPO}{proximal policy optimization}
 \acro{PRB}{physical resource block}
 \acro{RL}{reinforcement learning}
 \acro{RSRP}{reference signal received power}
 \acro{SAGIN}{space-air–ground-sea integrated network}
 \acro{SINR}{signal-to-interference-plus-noise ratio}
 \acro{TBS}{terrestrial base station}
 \acro{TD3}{Twin Delayed Deep Deterministic Policy Gradient}
 \acro{TN}{terrestrial network}
 \acro{UAV}{unmanned aerial vehicle}
 \acro{UAV-BS}{unmanned aerial vehicle-mounted base station}
 \acro{UE}{user equipment}
\end{acronym}

\begin{abstract}
Unmanned aerial vehicle (UAV)-mounted base stations are highly susceptible to wind disturbances such as gusts and turbulence, which induce positional drift and degrade communication link quality, particularly in emergency scenarios. To address this challenge, we propose a deep reinforcement learning-based framework for wind-resilient trajectory adjustment and positioning based on the Twin Delayed Deep Deterministic Policy Gradient (TD3) algorithm. The method models wind as a stochastic kinematic perturbation, avoiding complex aerodynamic modeling, thereby enabling the TD3 agent to learn adaptive control policies that maintain optimal coverage footprints.
By prioritizing user-centric performance metrics under turbulent conditions, the proposed architecture ensures continuous service availability despite external disruptions. Simulation results demonstrate that the TD3-based approach effectively compensates for wind-induced displacements and outperforms benchmark methods, including proximal policy optimization, in terms of throughput stability and robustness in windy environments.
\\
\end{abstract}


\section{Introduction} 

\Acp{UAV} are increasingly deployed as aerial base stations in wireless communication systems~\cite{ibanez2025optimizing}, offering flexible and on-demand connectivity, particularly in emergency scenarios. However, their performance is significantly degraded by environmental disturbances, especially wind and atmospheric turbulence. These disturbances induce trajectory deviations, instability, and reduced communication quality, thereby necessitating robust control and adaptation strategies~\cite{hasan2023evaluation}. Recent studies have highlighted the critical impact of atmospheric conditions on \ac{UAV-BS} systems. Surveys on wind-aware simulations and UAV-based atmospheric sensing further underscore the need for accurate turbulence modeling in realistic \ac{UAV-BS} deployments~\cite{sasse2023survey, cole1905spatio}.

\subsection{Related Work}


Substantial efforts have investigated \ac{UAV-BS} deployment and trajectory optimization from a communication perspective, targeting network metrics such as coverage, throughput, and fairness. However, most of these works assume ideal flight conditions and do not explicitly consider wind-induced disturbances.
At the same time, extensive research has addressed wind-effect modeling and robust disturbance rejection to enhance \ac{UAV-BS} trajectory performance in turbulent environments. 
Wind-resilient \ac{UAV-BS} trajectory control methods can be broadly categorized as follows:

\subsubsection{Classical Wind-Resilient UAV-BS Trajectory Control}

Significant efforts have been devoted to improving UAV-BSs stability and control under turbulent conditions. Traditional proportional-integral-derivative controllers, despite their simplicity, often exhibit degraded performance in the presence of nonlinear disturbances. In contrast, advanced model-based techniques such as cascaded model predictive control have demonstrated superior stability and disturbance rejection capabilities~\cite{sadi2025optimizing}. Pilot-in-the-loop simulations have been used to assess controllability limits and highlight the sensitivity of \ac{UAV}-BSs to atmospheric turbulence~\cite{hasan2023evaluation}.
Robust control methods, including sliding mode control methods~\cite{chen2022robust}, disturbance-rejection-based trajectory tracking~\cite{xu2023robust}, and fixed-time control~\cite{cai2024fixed}, have shown enhanced performance under uncertainties and external disturbances. Trajectory adaptation strategies such as dual-mode control further enable \acp{UAV} to adjust motion according to wind conditions while preserving control authority in extreme environments~\cite{cole2019trajectory}.
Nevertheless, these approaches generally rely on accurate system dynamics and predefined assumptions. Consequently, their effectiveness diminishes under highly dynamic and stochastic wind conditions. Moreover, their high computational complexity and limited adaptability hinder real-time applicability in \ac{UAV} trajectory optimization.

\subsubsection{DRL-Based Wind-Resilient UAV-BS Trajectory Control}



To address high computational complexity and limited adaptability under severe wind turbulence, \ac{DRL} has emerged as a promising paradigm for managing wind disturbances, aerodynamic uncertainties, and multi-objective trajectory optimization~\cite{lale2024falcon}.
For instance, \ac{DRL}-based \ac{UAV-BS} flight control using the \ac{TD3} and \ac{PPO} algorithms was investigated in~\cite{khanzada2025reinforcement}, where \ac{TD3} showed superior stability and faster convergence compared with \ac{PPO}. Focusing on \ac{TD3}, Himanshu \textit{et al.}~\cite{himanshu2022waypoint} reported improved navigation performance for quadrotors under disturbances, while~\cite{gamal2024control} achieved robust stabilization of a twin-rotor system in windy conditions.
Nevertheless, these studies primarily address low-level flight dynamics and navigation tasks. Existing works typically treat \ac{UAV-BS} control and communication performance independently, with limited integration of wind-resilient control strategies and communication-aware trajectory optimization in \ac{UAV-BS} networks.

\subsection{Motivation and Contributions}
Despite recent advances in wind-resilient \ac{UAV-BS} control, classical approaches~\cite{sadi2025optimizing, hasan2023evaluation, chen2022robust, xu2023robust, cai2024fixed, cole2019trajectory} rely on accurate system models and predefined dynamics, limiting their adaptability to highly dynamic and stochastic wind conditions. Although \ac{DRL}-based solutions~\cite{khanzada2025reinforcement, himanshu2022waypoint, gamal2024control} have shown promising results, most prior works focus on flight stability and navigation, with limited emphasis on communication-aware trajectory optimization. In contrast, our prior work~\cite{ibanez2025optimizing,akhtarshenas2026centralized} addressed UAV-BS positioning without accounting for wind effects or turbulence.
Motivated by this gap, particularly in emergency scenarios requiring reliable connectivity under rapidly changing conditions, user mobility, and incomplete system information, we propose a DRL-based framework for joint wind-resilient trajectory control and positioning.

The main contributions of this work are listed as follows:
\begin{itemize}
    \item 
    We develop a \ac{TD3}-based framework for adaptive UAV-BS positioning that captures UE mobility patterns, enhances coverage performance, and improves fairness-aware throughput in dynamic environments.
    \item 
   We enable UAV-BS operation in GPS-denied scenarios under realistic atmospheric conditions by exploiting UE reference signals and AoA measurements, providing robust positioning under turbulent conditions.
    \item 
    Simulation results demonstrate that the proposed TD3-based approach achieves faster convergence and outperforms \ac{PPO} in terms of mean reward and throughput.
\end{itemize}


\section{Atmospheric Wind Disturbance Modeling}

To evaluate the impact of atmospheric disturbances on \ac{UAV}-\ac{BS} performance, four wind conditions are considered: constant wind and wind shear (steady components), as well as discrete gust and continuous turbulence (stochastic components)~\cite{hasan2023evaluation}. Wind is modeled as the superposition of steady and stochastic components, representing persistent effects and turbulence, respectively~\cite{bohn2019deep}. In this work, these disturbances are incorporated as cumulative velocity perturbations on the \ac{UAV-BS} kinematic state, with emphasis on the networking-layer response (see Fig.~\ref{fig:wind_all}). Further details and mathematical formulations can be found in~\cite{phadke2024modeling,sadi2025optimizing,hasan2023evaluation,bohn2019deep}.
\subsubsection{Constant Wind}
As shown in Fig.~\ref{fig:wind_all}(a), constant wind is modeled as a steady disturbance that gradually increases to a specified value and remains constant thereafter. It is defined as $\mathbf{v}_{\text{const}}(t)=\mathbf{v}_0$, where $\mathbf{v}_0$ is a constant wind velocity vector, representing a persistent external force on \ac{UAV}-\ac{BS} motion~\cite{bohn2019deep}.

\subsubsection{Wind Shear}
Fig.~\ref{fig:wind_all}(b) illustrates wind shear, modeled as a gradual variation in wind speed (and possibly direction) over time. It is expressed as $\mathbf{v}_{\text{shear}}(t)=\mathbf{v}_0+\kappa t$, where $\kappa$ denotes the shear rate. This non-uniform wind profile induces deviations in the UAV-BS trajectory~\cite{hasan2023evaluation}.

\subsubsection{Discrete Gust}
A discrete gust is modeled as a short-duration disturbance with smooth rise and decay (Fig.~\ref{fig:wind_all}(c)):
\begin{equation}
\mathbf{v}_{\text{gust}}(t)=
\begin{cases}
V_g \sin\!\left(\dfrac{\pi (t-t_0)}{T_g}\right), & t_0 \le t \le t_0+T_g,\\
0, & \text{otherwise},
\end{cases}
\end{equation}
where $V_g$ is the peak gust amplitude and $T_g$ is the duration. This transient disturbance causes temporary deviations in \ac{UAV}-\ac{BS} motion~\cite{bohn2019deep, sadi2025optimizing, hasan2023evaluation}.

\subsubsection{Continuous Turbulence}
\begin{figure}[t]
\centering
\includegraphics[width=0.8\linewidth]{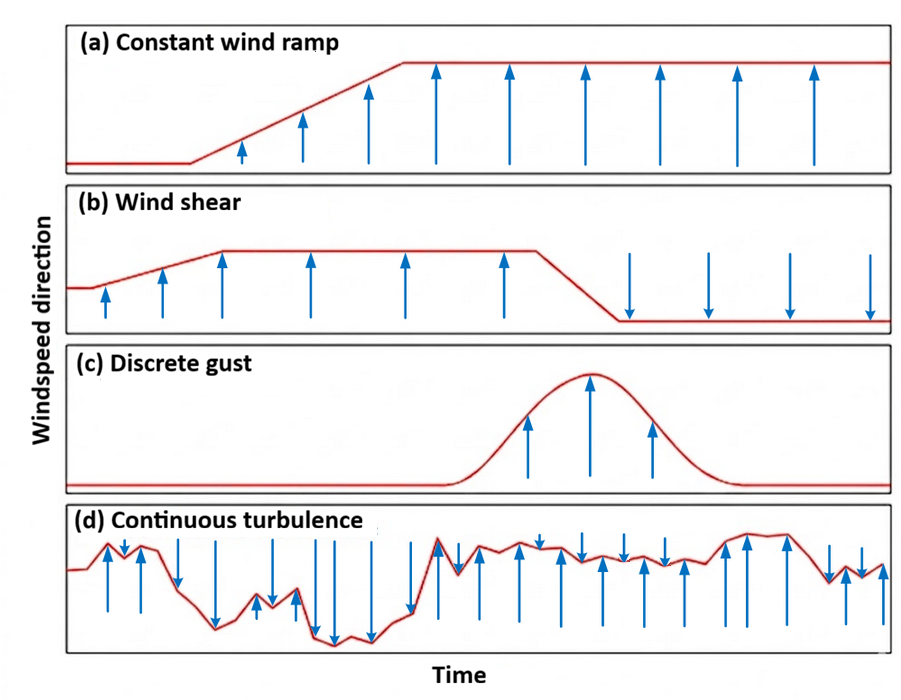}
\caption{Wind input profiles~\cite{hasan2023evaluation}.}
\label{fig:wind_all}
\end{figure}

Continuous turbulence is represented as a stochastic process with random fluctuations over time (Fig.~\ref{fig:wind_all}(d)). It is modeled using a first-order Gauss--Markov process~\cite{phadke2024modeling, hasan2023evaluation}: $\mathbf{v}_{\text{turb}}(t+1)=\alpha \mathbf{v}_{\text{turb}}(t)+\mathbf{w}(t)$, where $\alpha\in[0,1]$ determines temporal correlation and $\mathbf{w}(t)\sim\mathcal{N}(0,\sigma^2\mathbf{I})$ represents Gaussian white noise. This captures persistent and time-varying atmospheric effects on \ac{UAV}-\ac{BS}.

\section{System Description}
\label{sec:system_model}

In this study, we consider a wireless network composed of multiple \acp{UAV-BS} serving ground \acp{UE} (first responders).  
The system operates in both downlink and uplink using TDMA. Let $\mathcal{U} = \{u_1, \ldots, u_U\}$ denote the set of \acp{UE}. 
The position of UE $u \in \mathcal{U}$ is given by $\boldsymbol{\rho}_u^{\mathrm{U}} = [x_u^{\mathrm{U}}, y_u^{\mathrm{U}}, z_u^{\mathrm{U}}]^\top$, 
and the collective positions of all UEs are denoted by $\mathbf{X}^{\mathrm{U}} = [\boldsymbol{\rho}_1^{\mathrm{U}}, \ldots, \boldsymbol{\rho}_U^{\mathrm{U}}]$.
\acp{UE} are grouped into hotspots, 
where the communication metrics of all \ac{UE}s within each hotspot are aggregated and, when required, represented by a single equivalent \ac{UE}. 
Let $\mathcal{H} = \{h_1, \ldots, h_H\}$ denote the set of hotspots,
each with radius $r$ and equal number of \acp{UE}.
Hotspot mobility follows a linear mobility model. 
All \acp{UE} move with speed $v$. 
Each \ac{UE} in the same hotspot is associated with the \ac{UAV-BS} dedicated to this hotspot.
Let $\mathcal{D} = \{d_1, \ldots, d_D\}$ denote the set of \acp{UAV-BS}, 
with positions $\boldsymbol{\rho}^{\mathrm{D}}_d = [x^{\mathrm{D}}_d, y^{\mathrm{D}}_d, z^{\mathrm{D}}_d]^\top$. 
The collective \ac{UAV-BS} positions are given by $\mathbf{X}^{\mathrm{D}} = [\boldsymbol{\rho}^{\mathrm{D}}_1, \ldots, \boldsymbol{\rho}^{\mathrm{D}}_D]$.

\subsection{Channel Model}

In this study, we use the Urban Macro (UMa) channel models defined by the Third Generation Partnership Project (3GPP) in TR 36.814 to calculate each of the aforementioned channel gain components,
with the following modifications:
the multi-path fading is modeled using a Rician distribution, 
and the BS antennas are assumed to be omnidirectional.
We assume that our proposed network operates over a bandwidth $B$ at a carrier frequency $f$. 
Each radio link within the network is subject to both slow and fast fading effects.
The overall channel gain between the $u^{\text{th}}$ \ac{UE} belonging to any hotspot served by the $d^{\text{th}}$ \ac{UAV}-\acp{BS} on the frequency resource $k$ is denoted by $G_{u,d,k}$.
This composite gain is modeled as the product of several contributing factors: 
antenna gain ($G^{\rm a}$), path gain ($G^{\rm p}$), shadow fading ($G^{\rm s}$), and fast fading ($G^{\rm ff}$), as given by
\begin{equation}
    G_{u,d,k} = G^{\rm a}_{u,d} \cdot G^{\rm p}_{u,d} \cdot G^{\rm s}_{u,d} \cdot \left|G^{\rm ff}_{u,d,k}\right|^2.
\end{equation}
The overall channel gain, $G_{u,d,k}$, depends on both the position of each \ac{UAV-BS},
denoted by $\rho^\mathrm{D}_d$ (which is treated as an optimization variable), 
and the position of the UEs, denoted by $\rho^\mathrm{U}_u$, 
which is assumed to be unknown. 
Therefore, the overall channel gain between the $u^{\text{th}}$ UE and the $d^{\text{th}}$ \ac{UAV-BS} can be expressed as $G_{u,d,k} = G_{u,d,k}\big(\boldsymbol{\rho}^\mathrm{U}_u, \boldsymbol{\rho}^\mathrm{D}_d\big)$.
The power received by the $u^{\text{th}}$ UE from the $d^{\text{th}}$ \ac{UAV-BS} on the the $k^{\text{th}}$ frequency resource is given by
\begin{equation}
    P^{\rm rx}_{u,d,k}(\boldsymbol{\rho}^{\mathrm{U}}_u, \boldsymbol{\rho}^{\mathrm{D}}_d)
    = P^{\rm tx}_{d,k} \cdot G_{u,d,k}(\boldsymbol{\rho}^{\mathrm{U}}_u, \boldsymbol{\rho}^{\mathrm{D}}_d).
    \label{eq_intro:Received_power}
\end{equation}
where $P^{\rm tx}_{d,k}$ denotes the transmit power of the the $d^{\text{th}}$ \ac{UAV-BS} on the $k^{\text{th}}$ frequency resource.

The signal quality experienced by the $u^{\text{th}}$ UE from the $d^{\text{th}}$ \ac{UAV-BS} on the frequency resource $k$ is quantified by the average \ac{SINR}, denoted by $\gamma_{u,d,k}$. 
It is computed as~\cite{goldsmith2005wireless}
\begin{equation}
    \gamma_{u,d,k}(\boldsymbol{\rho}^{\mathrm{U}}_u, \bm{\mathrm{X}}^{\mathrm{D}})
    =
    \frac{P^{\mathrm{rx}}_{u,d,k}(\boldsymbol{\rho}^{\mathrm{U}}_u, \boldsymbol{\rho}^{\mathrm{D}}_d)}
    {\sum_{\substack{d' = 1 \\ d' \neq d}}^{D}
    P^{\mathrm{rx}}_{u,d',k}(\bm{\rho}^{\mathrm{U}}_u, \bm{\rho}^{\mathrm{D}}_{d'}) + \sigma^2_k}.
    \label{eq_intro:sinr}
\end{equation}
where $\sigma^2_k$ denotes the noise power in the frequency resource $k$, and $\bm{\rho}^{\mathrm{D}}$ represents the set of positions of all \acp{UAV-BS}.
Using Shannon--Hartley theorem,
the achievable throughput for the first responder $u$ connected to the \ac{BS} of the \ac{UAV-BS} $d$ on the frequency resource $k$ can be expressed as~\cite{goldsmith2005wireless}
\begin{equation}
    R_{u,d,k}(\bm{\rho}^{\mathrm{U}}_u, \bm{\mathrm{X}}^{\mathrm{D}})
    =
    B_k \log_2 \left(1 + \gamma_{u,d,k}(\bm{\rho}^{\mathrm{U}}_u, \bm{\mathrm{X}}^{\mathrm{D}})\right).
    \label{eq_intro:rate}
\end{equation}
where $B_k$ denotes the bandwidth allocated to the frequency resource $k$. 
If a scheduler is used to ensure fair distribution of the available resources among the UEs inside the cell—such as a round-robin scheme-the achievable throughput of the $u^{\text{th}}$ \ac{UE} associated with the $d^{\text{th}}$ \ac{UAV-BS} can be represented as
\begin{equation}
    R_{u,d}(\bm{\rho}^{\mathrm{U}}_u, \bm{\mathrm{X}}^{\mathrm{D}})
    =
    \frac{B}{U} \log_2 \left(1 + \bar{\gamma}_{u,d}(\bm{\rho}^{\mathrm{U}}_u, \bm{\mathrm{X}}^{\mathrm{D}})\right).
    \label{eq_intro:mean_rate}
\end{equation}
where $\frac{B}{U}$ denotes the average bandwidth allocated per \ac{UE} under equal resource sharing, 
and $\bar{\gamma}_{u,d}(\bm{\rho}^{\mathrm{U}}_u, \bm{\mathrm{X}}^{\mathrm{D}})$ represents the effective \ac{SINR} averaged over the allocated frequency resources.


\section{DRL-based Problem Statement}
Considering the proposed system model in the previous section, a \ac{DRL}-based framework for a \ac{UAV-BS} trajectory optimization under wind turbulence is considered, 
to optimize network performance using a fairness-based throughput metric. 

\subsection{Objective Function}

The goal is to determine, in real time, the optimal \acp{UAV-BS} positions $\bm{\rho}^\mathrm{D}(t)$ that maximize the total fair throughput, denoted by \( R_{\text{fair}} \), for the \acp{UAV-BS} under wind turbulence. To this end, a centralized architecture is adopted, where a single \ac{RL} agent controls the \acp{UAV-BS} to provide communication services.
We analyze the system over a finite \ac{UE} movement and \ac{UAV-BS} operation period $T$, 
indexed by discrete time steps $t \in \{0,1,\ldots,T\}$. 
For example, at time step $t$, the location of the $u$-th \ac{UE} is denoted by $\boldsymbol{\rho}^{\mathrm{U}}_{u}(t)$, 
while the location of the $d$-th \ac{UAV-BS} is given by $\boldsymbol{\rho}^{\mathrm{D}}_{d}(t)$. 
For notational simplicity, the explicit time index \(t\) is omitted in the remainder of this paper.
The total fair throughput is defined as
\begin{equation}
R_{\mathrm{fair}}(\mathbf{X}^{\mathrm{U}}, \mathbf{X}^{\mathrm{D}})
=
\sum_{d=1}^{D}
\sum_{u \in \mathcal{U}_d}
\log_{10}
\left(
R_{u,d}
\left(
\bm{\rho}^{\mathrm{U}}_u, \bm{\mathrm{X}}^{\mathrm{D}}
\right)
\right),
\label{eq_intro:fair_rate}
\end{equation}
where $\mathbf{X}^{\mathrm{U}}$ and $\mathbf{X}^{\mathrm{D}}$ denote the (unknown) locations of all UEs and the positions of UAV-BSs, respectively. $\mathcal{U}_d$ denotes the set of UEs associated with the $d^{\text{th}}$ UAV-BS, with $|\mathcal{U}_d| = N$ in the considered setup. 
This formulation promotes network fairness by giving higher priority to \acp{UE} with lower throughput,
thus avoiding resource allocation bias toward already well-served \acp{UE}.
The optimization problem is given by:
\begin{equation}
    \max_{\mathbf{X}^\mathrm{D}} \; R_{\mathrm{fair}}(\mathbf{X}^{\mathrm{U}}, \mathbf{X}^{\mathrm{D}}),
\end{equation}
where $\mathbf{X}^{\mathrm{D}}$ denotes the positions of the UAV-BS to be optimized.
Real-time UAV-BS trajectory optimization is challenging due to the high dimensionality, stochasticity, nonlinearity, wind turbulence, UE mobility, and dynamic radio conditions. Therefore, we employ \ac{DRL}, specifically TD3, which provides stable and sample-efficient learning for adaptive real-time decision-making in continuous control environments.

\subsection{DRL-based TD3}

The \ac{UAV-BS} trajectory optimization problem under wind turbulence is formulated as a Markov decision process. 
The agent interacts with the environment through states $s \in \mathcal{S}$, actions $a \in \mathcal{A}$, and rewards $r$.
Transitions are recorded at each time step as $(s, a, r, s')$, 
where $s'$ denotes the next state. 
The objective of the agent is to learn a policy $\pi$ that maximizes the expected long-term return (see Algorithm~\ref{Algorithmtd3}).
\Ac{TD3}~\cite{fujimoto2018addressing} is an off-policy actor-critic algorithm that improves training stability and reduce overestimation bias in continuous control tasks. 
\ac{TD3} learns a deterministic policy $\pi_{\phi}(s)$ and two critic networks $Q_{\theta_1}(s,a)$ and $Q_{\theta_2}(s,a)$ using a replay buffer $\mathcal{B}$, enabling sample-efficient learning in stochastic environments.
To improve training stability, a warm-up phase of $N_{\text{warm}}$ steps is employed, 
during which actions are randomly sampled to populate the replay buffer before policy updates begin. 
The target for the critic networks is defined as~\cite{fujimoto2018addressing}:
\begin{equation}
    y = r + \gamma \min_{i=1,2} Q_{\theta_i'}(s', \tilde{a}),
\end{equation}
where $\tilde{a} = \pi_{\phi'}(s') + \epsilon$,
and $\epsilon \sim \text{clip}(\mathcal{N}(0,\sigma), -c, c)$.
The critic networks are trained by minimizing the loss function:
\begin{equation}
    \mathcal{L}(\theta_i) = \mathbb{E}_{\mathcal{B}} \left[ \left( Q_{\theta_i}(s,a) - y \right)^2 \right].
\end{equation}

The actor is updated using the deterministic policy gradient:
\begin{equation}
    \nabla_{\phi} J(\phi) = \mathbb{E}_{\mathcal{B}} \left[ \nabla_a Q_{\theta_1}(s,a)\big|_{a=\pi_{\phi}(s)} \nabla_{\phi} \pi_{\phi}(s) \right].
\end{equation}

\ac{TD3} further improves stability through clipped double Q-learning, target policy smoothing, and delayed policy updates, 
which reduce value overestimation and training variance.
For the $\tau \ll 1$, the target networks are softly updated as~\cite{fujimoto2018addressing}:
\begin{equation}
    \theta_i' \leftarrow \tau \theta_i + (1 - \tau)\theta_i', \quad
    \phi' \leftarrow \tau \phi + (1 - \tau)\phi',
\end{equation}

In contrast to on-policy \ac{PPO}, 
\ac{TD3} leverages experience replay to improve sample efficiency and robustness. 
This makes \ac{TD3} suitable for dynamic and non-stationary environments, 
such as wind-affected \ac{UAV-BS} operations.
\begin{algorithm}
\caption{TD3~\cite{fujimoto2018addressing}}
\begin{algorithmic}[1]
\State Initialize critic networks $Q_{\theta_1}, Q_{\theta_2}$ and actor network $\pi_{\phi}$ with random parameters $\theta_1, \theta_2, \phi$
\State Initialize target networks $\theta_1' \leftarrow \theta_1$, $\theta_2' \leftarrow \theta_2$, $\phi' \leftarrow \phi$
\State Initialize replay buffer $\mathcal{B}$
\For{$t = 1$ to $T$}
    \State Select action $a \sim \pi_{\phi}(s) + \epsilon$, $\epsilon \sim \mathcal{N}(0,\sigma)$
    \State Observe reward $r$ and next state $s'$
    \State Store $(s,a,r,s')$ in $\mathcal{B}$
    \State Sample mini-batch of $N$ transitions $(s,a,r,s')$ from $\mathcal{B}$
    \State $\tilde{a} \leftarrow \pi_{\phi'}(s') + \epsilon$, \quad $\epsilon \sim \text{clip}(\mathcal{N}(0,\sigma), -c, c)$
    \State $y \leftarrow r + \gamma \min_{i=1,2} Q_{\theta_i'}(s', \tilde{a})$
    \State Update critics:
    \[
    \theta_i \leftarrow \arg\min_{\theta_i} \frac{1}{N} \sum (y - Q_{\theta_i}(s,a))^2
    \]
    \If{$t \bmod d = 0$}, Update actor:
        \[
        \nabla_{\phi} J(\phi) = \frac{1}{N} \sum \nabla_a Q_{\theta_1}(s,a)\big|_{a=\pi_{\phi}(s)} \nabla_{\phi} \pi_{\phi}(s)
        \]
        \State Update target networks:
        \[
        \theta_i' \leftarrow \tau \theta_i + (1-\tau)\theta_i'
        \]
        \[
        \phi' \leftarrow \tau \phi + (1-\tau)\phi'
        \]
    \EndIf
\EndFor
\end{algorithmic}
\label{Algorithmtd3}
\end{algorithm}

\subsubsection{State Space}

Here, we introduce the \ac{DRL} agent's state space for the proposed system model. 
During each episode of length $\tau$ time steps, the state set comprises time-dependent observations from all \acp{UAV-BS}, including their positions, the average \ac{SINR}, and AoA statistics of each hotspot measured by the on-board \ac{BS} \cite{barrios2026passive}. Specifically, the state of the $d^{\text{th}}$ \ac{UAV-BS} associated with the $h^{\text{th}}$ hotspot is defined as:
\begin{equation}
\mathcal{S}
=
\left\{
\tilde{\boldsymbol{\rho}}_{d},
\bar{\gamma}_{d},
\Psi_{d,h},
\Phi_{d,h}
\right\}_{d=1}^{D}.
\label{eq:state_space}
\end{equation}
where $\tilde{\boldsymbol{\rho}}_d = [\boldsymbol{\rho}^{\mathrm{D}}_{d,t}, \ldots, \boldsymbol{\rho}^{\mathrm{D}}_{d,t-\tau}]$ denotes the position history of the \ac{UAV-BS} $d$ over a window of $\tau$ time steps. At time step $t=i$, $\boldsymbol{\rho}_{d,i}$ represents the position of the $d^{\text{th}}$ UAV-BS, while $\bar{\gamma}_{d,i}$ denotes the average \ac{SINR} of all \acp{UE} served by UAV-BS~$d$. 
Furthermore, $\Psi_{d,h,i}$ and $\Phi_{d,h,i}$ represent the circular mean and circular standard deviation of the AoA measurements associated with hotspot $h$, respectively. These AoA-based statistics encode directional characteristics of the \ac{UE} distribution, allowing the agent to infer spatial user dynamics without requiring explicit \ac{UE} location information. 
Assuming three UAV-BSs operating at a fixed altitude and serving three hotspots, the overall state space is defined as:
\begingroup
\setlength{\abovedisplayskip}{6pt}
\setlength{\belowdisplayskip}{6pt}

\begin{equation}
\label{eq:observation_set}
\mathcal{S} =
\left\{
\begin{aligned}
&\big((x_1,y_1,h_{\mathrm{UAV}}),\, \bar{\gamma}_{1,1},\, \Psi_{1,1},\, \Phi_{1,1}\big),\\
&\big((x_2,y_2,h_{\mathrm{UAV}}),\, \bar{\gamma}_{2,2},\, \Psi_{2,2},\, \Phi_{2,2}\big),\\
&\big((x_3,y_3,h_{\mathrm{UAV}}),\, \bar{\gamma}_{3,3},\, \Psi_{3,3},\, \Phi_{3,3}\big).
\end{aligned}
\right\}.
\end{equation}
\endgroup

\subsubsection{Action Space}

At every decision stage, 
the action specifies the movement of each \ac{UAV-BS}.
The agent selects a direction $\beta_d$ and a displacement magnitude $r$ within a continuous action space,
i.e., $\mathcal{A} = \{(\beta_d, r)\}$, 
where $\beta_d \in [-180^\circ, 180^\circ)$, $r \in [0, r_{\max}]$, and $d \in \{1,2,3\}$.
The parameter $r_{\max}$ defines the maximum allowable displacement per time step, 
and $\beta_d$ is measured with respect to the east direction.
The action is defined for each \ac{UAV-BS}, 
and the overall action space consists of the joint movement decisions of all \acp{UAV-BS}. 
The proposed framework leverages the continuous control capability of \ac{TD3} to enable fine-grained trajectory adjustments without action discretization, 
which is essential for capturing small spatial variations that significantly impact network performance.

\subsubsection{Reward Function}

The reward is designed to maximize the total fair throughput $R_{\mathrm{fair}}$,
as defined in \eqref{eq_intro:fair_rate}.
To ensure numerical stability and consistent reward scaling, 
a min-max normalization is applied to the reward signal.
To further stabilize training,
a parametric nonlinear mapping based on a sigmoid function is employed,
transforming $R_{\mathrm{fair}}$ as:
\begin{equation}
\label{eq:sigmoid}
    \mathcal{R} = \left(1 + \exp\left(-c_s (R_{\text{fair}} - c_m)\right)\right)^{-1}.
\end{equation}
where $c_s > 0$ controls the slope and $c_m$ determines the midpoint of the mapping, set to $c_s = 0.25$ and $c_m = 20$, respectively. The parameters are tuned using two representative scenarios: static UAV-BSs above hotspots and dynamically moving UAV-BSs. Average episode throughput over multiple realizations is used to select $c_s$ and $c_m$, ensuring appropriate reward scaling and learning sensitivity.




\begin{figure*}[!htb]
\centering
\begin{subfigure}{0.433\textwidth}
\centering
\includegraphics[width=\linewidth]{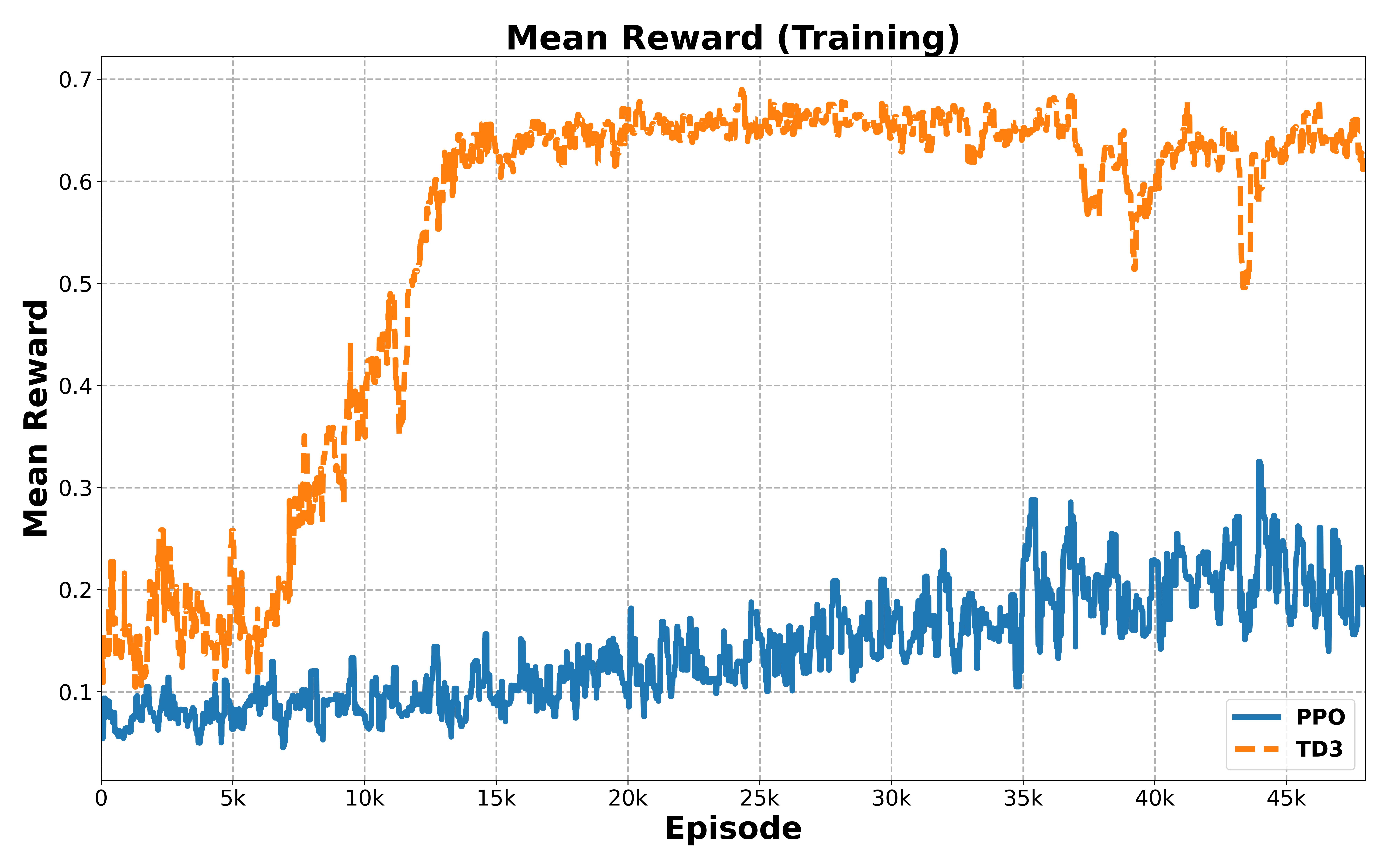}
\caption{Training reward (TD3 vs. PPO). }
\label{fig1}
\end{subfigure}
\hfill
\begin{subfigure}{0.433\textwidth}
\centering
\includegraphics[width=\linewidth]{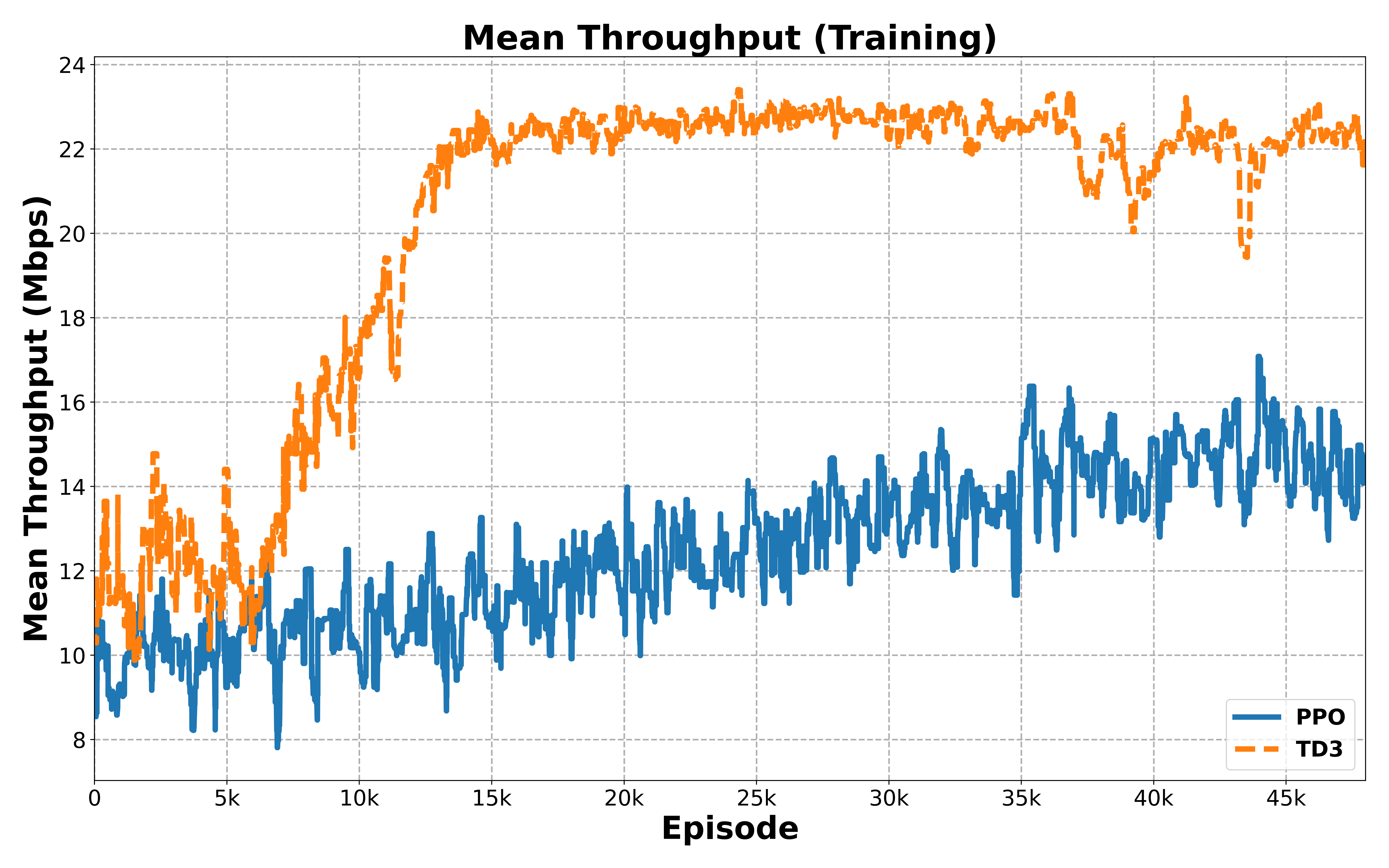}
\caption{Training throughput (TD3 vs. PPO).}
\label{fig2}
\end{subfigure}
\vspace{0.3cm}

\begin{subfigure}{0.433\textwidth}
\centering
\includegraphics[width=\linewidth]{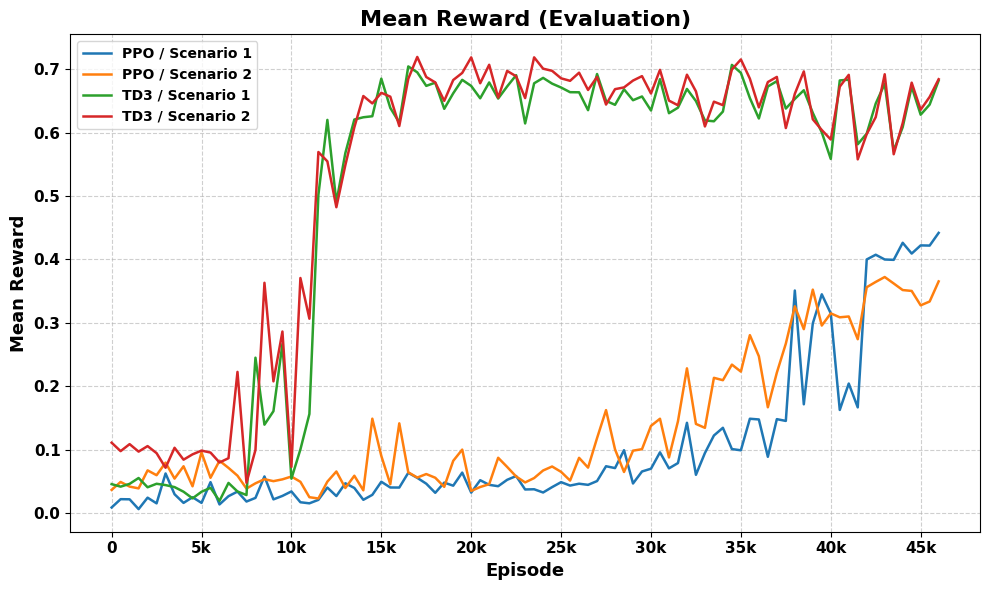}
\caption{Evaluation reward (TD3 vs. PPO).}
\label{fig3}
\end{subfigure}
\hfill
\begin{subfigure}{0.433\textwidth}
\centering
\includegraphics[width=\linewidth]{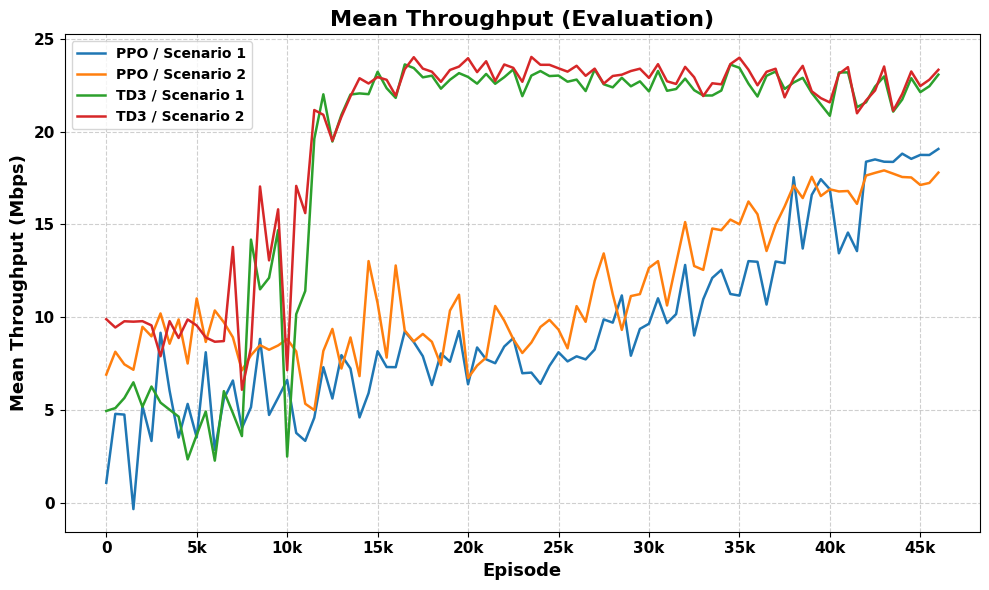}
\caption{Evaluation throughput (TD3 vs. PPO).}
\label{fig4}
\end{subfigure}

\caption{Trajectory optimization: comparison of TD3 and PPO performance during training and evaluation under wind turbulence.}
\label{fig:combined_results}
\end{figure*}

\section{Results and Analysis}


The simulation setup is described as follows:
\subsubsection*{Network Settings}

We consider a rectangular deployment area of \SI{400}{m} $\times$ \SI{1200}{m}, symmetric about the origin, with three hotspots initially located at $(-350, -550)$, $(350, 0)$, and $(-350, 550)$ (all coordinates in meters), following linear \ac{UE} mobility. Each hotspot serves 10 \acp{UE}, deployed at a fixed altitude of \SI{1.5}{m} and moving at a constant speed of \SI{10}{m/s}. The hotspots move along straight lines from right to left and reverse direction upon reaching the deployment boundaries.
The environment includes three \acp{UAV-BS} operating at an altitude of \SI{100}{m},
each with a mass of \SI{2.5}{kg} and an initial speed of \SI{10}{m/s}.
The maximum displacement per time step is limited to $r_{\max} = \SI{20}{m}$.
Each \ac{UAV-BS} operates with a system bandwidth of \SI{5}{MHz}, a carrier frequency of \SI{2}{GHz}, a transmit power of \SI{43}{dBm}, and a noise power is \SI{-114}{dBm/MHz}. The propagation scenario is based on the 3GPP UMa model (TR 36.814), with omnidirectional antennas and Rician fading.
\subsubsection*{Wind Disturbance Settings}
Low-altitude wind is modeled as a composite process including constant, stochastic, shear, and gust components. 
The steady component has an average speed of \SI{6.5}{m/s}, representing typical operating conditions.
Stochastic turbulence is modeled using an autoregressive AR(1) process with a standard deviation of \SI{1.5}{m/s},
capturing short-term fluctuations.
A slowly varying shear component with amplitude \SI{1.0}{m/s} and a period of 180 steps accounts for gradual atmospheric variations over time.
In addition, a transient gust event with a peak speed of \SI{2.5}{m/s} is introduced at step 80 for a duration of 8 steps, 
modeling wind disturbances~\cite{bohn2019deep, hasan2023evaluation}.

\subsubsection*{Neural Network Settings}

The main \ac{TD3} hyperparameters are defined as follows. 
Both the actor and critic networks use a constant learning rate of $3 \times 10^{-4}$ and consist of three hidden layers with 256 neurons each.
The \ac{TD3} algorithm uses a replay buffer of size $10^6$, a batch size of 256, a discount factor $\gamma = 0.99$, and a soft update coefficient $\tau = 0.005$.
Gradient clipping with a maximum norm of 1.0 is applied to stabilize training.
In addition, delayed policy updates with a factor of 2, target policy smoothing with clipped noise ($\sigma = 0.2$, clip $= 0.5$), and decaying exploration noise (initial standard deviation $= 0.2$) are employed. 
For \ac{PPO}, a constant learning rate of $\eta = 3 \times 10^{-5}$ is used, with both actor and critic networks having the same architecture of three hidden layers with 256 neurons each.
\ac{PPO} uses $\gamma = 0.99$ and $\lambda = 0.95$ to balance short- and long-term rewards. 
For both algorithms, training is performed over 46k episodes, consisting of 256 time-steps.

\subsubsection*{Numerical Results} To assess generalization, the performance of the proposed system is evaluated under two scenarios. During training, the initial \acp{UAV-BS} positions are randomly initialized, and the reward and throughput are monitored accordingly (See Fig.~\ref{fig:combined_results}(a) and (b)). During evaluation, the \acp{UAV-BS} are initialized with two predefined positions at the altitude of 100\,m and horizontal coordinates $[(0,180),(0,0),(0,-180)]$ for the first scenario, and $[(-180,380),(180,380),(-180,-380)]$ for the second scenario correspondingly. The system performance is assessed every 500 iterations (See Fig.~\ref{fig:combined_results}(c) and (d)).
Fig.~\ref{fig:combined_results} presents the simulation results evaluating both communication performance (throughput) and \ac{RL} performance (reward). The figure compares off-policy \ac{TD3} and on-policy \ac{PPO} under wind turbulence during the training phase (Fig.~\ref{fig:combined_results}(a) and (b)) and the evaluation phase (Fig.~\ref{fig:combined_results}(c) and (d)).
During training, Fig.~\ref{fig:combined_results}(a) shows that \ac{TD3} achieves a higher and more stable mean reward, converging to approximately 0.65, while \ac{PPO} converges to a lower value with noticeable fluctuations. Fig.~\ref{fig:combined_results}(b) further demonstrates that \ac{TD3} rapidly reaches a mean throughput of approximately 22 Mbps within the early training stages (around $10^4$ episodes), whereas \ac{PPO} improves more gradually and exhibits higher variance.
As shown in Fig.~\ref{fig:combined_results}(c), \ac{TD3} consistently achieves higher reward values with lower variability, indicating more effective optimization of the fairness-based objective. Fig.~\ref{fig:combined_results}(d) confirms that \ac{TD3} maintains a mean throughput of approximately 22 Mbps across different initializations, while \ac{PPO} converges to approximately 16 Mbps with increased variability.
These results demonstrate that \ac{TD3} provides more robust and consistent performance under dynamic wind conditions.

\section{Conclusion}
This paper presented a \ac{TD3}-based \ac{DRL} framework for wind-resilient \ac{UAV-BS} trajectory optimization. By modeling wind as a stochastic disturbance and incorporating communication-aware objectives, the proposed approach enables adaptive \ac{UAV-BS} positioning while maintaining service continuity.
Simulation results demonstrate that \ac{TD3} consistently outperforms \ac{PPO} in terms of convergence speed, stability, and communication performance, achieving higher reward and throughput under varying wind conditions. These findings highlight the effectiveness of \ac{TD3} in handling dynamic and non-stationary environments.
Overall, the proposed framework provides a practical and robust solution for \ac{UAV-BS}-assisted emergency communications, particularly in challenging scenarios such as wind disturbances and GPS-denied environments.

\vspace{12pt}

\bibliography{bibliography}

\end{document}